\documentclass[twocolumn,groupedaddress,showpacs,nofootinbib]{revtex4-1}

\usepackage[colorlinks,pdfstartview=FitH]{hyperref}
\hypersetup{linkcolor=blue,citecolor=blue,filecolor=black,urlcolor=blue}

\usepackage{amsmath,amssymb,bm}
\usepackage{graphicx}
\usepackage{amsfonts}
\usepackage{color}

\begin{document}


\baselineskip=15pt \parskip=5pt

\vspace*{3em}

\title{Velocity-dependent self-interacting dark matter from thermal freeze-out and tests in direct detections}

\author{Lian-Bao Jia}
\email{jialb@mail.nankai.edu.cn}
\affiliation{School of Science, Southwest University of Science and Technology, Mianyang
621010, China}

\begin{abstract}

A small fraction of millicharged dark matter (DM) is considered in the literature to give an interpretation about the enhanced 21-cm absorption at the cosmic dawn. Here we focus on the case that the main component of DM is self-interacting dark matter (SIDM), motivated by the small scale problems. For self interactions of SIDM being compatible from dwarf to cluster scales, velocity-dependent self interactions mediated by a light scalar $\phi$ is considered. To fermionic SIDM $\Psi$, the main annihilation mode $\Psi \bar{\Psi} \to \phi \phi$ is a $p -$wave process. The thermal transition of SIDM $\rightleftarrows \phi \rightleftarrows$ standard model (SM) particles in the early universe sets a lower bound on couplings of $\phi$ to SM particles, which has been excluded by DM direct detections, and here we consider SIDM in the thermal equilibrium via millicharged DM. For $m_\phi >$ twice millicharged DM mass, $\phi$ could decay quickly and avoid excess energy injection to the big bang nucleosynthesis. Thus, the $\phi -$SM particle couplings could be very tiny and evade DM direct detections. The picture of weakly interacting massive particle (WIMP)-nucleus scattering with contact interactions fails for SIDM-nucleus scattering with a light mediator, and a method is explored in this paper, with which a WIMP search result can be converted into the hunt for SIDM in direct detections.

\end{abstract}

\maketitle


\section{Introduction}

Modern astronomical observations \cite{Aghanim:2018eyx} indicate that dark matter (DM) accounts for about 84\% of the matter density in our universe, while the particle characters of DM, e.g., masses, components and interactions, etc, are currently unclear yet. If DM and ordinary matter are in thermal equilibrium in the very early universe, the DM particles would be thermal freeze-out with the expansion of the universe. One of the popular thermal freeze-out DM candidates is weakly interacting massive particles (WIMPs) with masses in a range of GeV$-$TeV scale. For WIMP type DM, the target nucleus could acquire a large recoil energy in WIMP-nucleus scattering in DM direct detections. Yet, confident WIMP signals are still absent from recent sensitive direct detections \cite{Liu:2019kzq,Abdelhameed:2019hmk,Agnese:2017jvy,Agnes:2018ves,Cui:2017nnn,Akerib:2016vxi,Akerib:2018hck,Aprile:2018dbl,Akerib:2017kat,Xia:2018qgs,Aprile:2019dbj,Amole:2019fdf}.

DM may have multi-components. Recently, a strong than expected 21-cm absorption at the cosmic dawn was reported by EDGES \cite{Bowman:2018yin}, and a possible explanation is that neutral hydrogen was cooled by the scattering with a small fraction of MeV millicharged DM \cite{Munoz:2018pzp,Fialkov:2018xre,Barkana:2018lgd,Barkana:2018cct,Berlin:2018sjs,Slatyer:2018aqg,Liu:2018uzy,Munoz:2018jwq,Jia:2018csj,Mahdawi:2018euy,Kovetz:2018zan,Jia:2019yhr,Liu:2019knx}. If so, what is the main component of DM? In addition, the $\Lambda$CDM model is successful in explaining the large-scale structure of the Universe, while deviations appear in small scales ($\lesssim$ 10 kpc), such as the core-cusp problem, missing satellites problem, and too-big-to-fail problem, etc (see e.g., Refs. \cite{Weinberg:2013aya,Bull:2015stt,Tulin:2017ara,Bullock:2017xww} for more). These small-scale problems may indicate some characters about the main component of DM, and possible strong self-interactions between DM
particles could provide a solution to the core-cusp and too-big-to-fail problems \cite{Spergel:1999mh,Vogelsberger:2012ku,Zavala:2012us,Foot:2014uba,Foot:2016wvj,Kaplinghat:2015aga,Tulin:2017ara,Kamada:2016euw,Valli:2017ktb}.\footnote{See Refs. \cite{Dodelson:1993je,Colombi:1995ze,Bode:2000gq,Lovell:2013ola} for the scenario of warm DM to the small-scale problems.} In this paper, the main component of DM is considered to be self-interacting dark matter (SIDM).

For collisional SIDM, to resolve the small-scale problems, the required scattering cross section per unit DM mass $\sigma/m_{\mathrm{DM}}$ is $\gtrsim$ 1 cm$^2$/g, while constraints from cluster collisions indicate that $\sigma/m_{\mathrm{DM}}$ should be $\lesssim$ 0.47 cm$^2$/g \cite{Randall:2007ph,Harvey:2015hha} (see Ref. \cite{Tulin:2017ara} for a recent review). In addition, the density profiles of galaxy clusters indicate that the corresponding self-interaction should be $\lesssim$ 0.1$-$0.39 cm$^2$/g \cite{Kaplinghat:2015aga,Harvey:2018uwf,Elbert:2016dbb}. This tension could be relaxed if the scattering cross section of SIDM is velocity dependent. Here we consider the light mediator being a scalar $\phi$, which couples to the Standard Model (SM) sector via the Higgs portal. When the mass of the mediator $m_\phi$ is much smaller than the SIDM mass (outside the Born limit), the scattering could be enhanced at low velocities \cite{Feng:2009hw,Tulin:2013teo}. Thus, the self interactions of SIDM could be compatible from dwarf to cluster scales.

For fermionic SIDM $\Psi$, the annihilation $\Psi \bar{\Psi} \to \phi \phi$ is a $p -$wave process. In the early universe, if SIDM and the SM particles were in the thermal equilibrium for a while via the transitions SIDM $\rightleftarrows \phi \rightleftarrows$ SM particles, this thermal equilibrium sets a lower bound on the couplings of $\phi$ to SM particles \cite{Chu:2011be,Dolan:2014ska,Jia:2016pbe}. For the light $\phi$ required by the velocity-dependent scattering between SIDM particles, the lower bound of the $\phi -$SM particle couplings set by the thermal equilibrium has been excluded by the present DM direct detections \cite{Jia:2016pbe}.\footnote{For example, for the case of the SIDM mass $\sim$ 20 GeV and [$m_\phi$/SIDM mass] $\sim 10^{-2}$, the SIDM-nucleon scattering cross section set by the thermal equilibrium is $\gtrsim$ 10$^{-40}$ cm$^2$, which has been excluded by DM direct detection experiments.} Thus, this type thermal freeze-out SIDM has been excluded by direct detections, and freeze-in SIDM is considered in the literature \cite{Duch:2017khv,Zakeri:2018hhe,Hambye:2018dpi}.

For velocity-dependent SIDM required to solve the small-scale problems, if the relic abundance of SIDM was set by the thermal freeze-out mechanism in the early universe, how to evade present constraints becomes an issue (especially DM direct detections). This is of our concern in this paper. For multi-component DM, besides the thermal equilibrium via SIDM $\rightleftarrows \phi \rightleftarrows$ SM particles, SIDM could be in the thermal equilibrium with  millicharged DM, which was in the thermal equilibrium with SM particles in the early universe and could give an explanation about the anomaly 21-cm absorption at the cosmic dawn. To avoid the excess energy injection into the period of the big bang nucleosynthesis (BBN) or an overabundance of $\phi$, the lifetime of $\phi$ should be much smaller than 1 second, and this can be achieved in the case of $m_\phi >$ twice millicharged DM mass. Thus, SIDM could be in the thermal equilibrium with SM particles via millicharged DM, and the $\phi -$SM particle couplings could be very tiny and evade DM direct detections. In addition, for SIDM-target nucleus scattering mediated by a light mediator, the momentum transfer could be comparable with the mediator mass $m_\phi$ in direct detections, and SIDM-nucleus scatterings would be different from WIMP-nucleus scatterings \cite{Kaplinghat:2013yxa}. The scenario above will be explored in this paper.

The following of this paper is organized as follows. The interactions in the new sector will be presented, and the self interactions of SIDM will be discussed in the next. Then, the direct detection of SIDM will be elaborated. The last part is the conclusion.

\section{Interactions in the new sector}

In this paper, two possible components of DM, the main component of SIDM $\Psi$ and a small fraction of millicharged DM $\chi$, are of our concern. For a small fraction of millicharged DM, it could give an explanation about the 21-cm absorption, and possible interactions between millicharged DM and SM particles have been studied in Refs. \cite{Berlin:2018sjs,Jia:2018csj,Jia:2019yhr}. Here we focus on SIDM, i.e., key transitions or interactions between SIDM and millicharged DM, SM particles. The effective interactions mediated by a new scalar field $\Phi$ are
\begin{eqnarray}
\mathcal{L}_i  &=& - \lambda \Phi \bar{\Psi} \Psi - \lambda_0 \Phi \bar{\chi} \chi - \mu_h \Phi (H^\dag H -\frac{V^2}{2}) \nonumber \\
 && - \lambda_h \Phi^2 (H^\dag H -\frac{V^2}{2})  - \frac{ \mu}{3!} \Phi^3  - \frac{ \lambda_4}{4!} \Phi^4   ~,
\end{eqnarray}
where $V$ is the vacuum expectation value, with $V \approx$ 246 GeV. The $\Phi$ field mixes with the Higss field after the electroweak symmetry breaking, and a mass eigenstate $\phi$ is generated (see e.g., Ref. \cite{Jia:2017kjw}). Here we suppose the mixing is very tiny, and thus $\phi$'s couplings to $\Psi$ and $\chi$ can be taken as equal to that of the corresponding $\Phi$'s couplings. The effective couplings of $\phi$ to SM fermions can be written as
\begin{eqnarray}
\mathcal{L}^i_{\phi f} = - \theta_\mathrm{mix} \frac{m_f}{V} \phi \bar{f} f,
\end{eqnarray}
where the mixing parameter $\theta_\mathrm{mix}$ is very tiny compared with 1. Here the particles playing important roles in transitions between DM and SM sectors are of our conern. There may be more particles in the new sector, and DM particles may also be composite particles \cite{Bhattacharya:2013kma,Cline:2013zca,Laha:2013gva,Laha:2015yoa,Kribs:2016cew,Kopp:2016yji,Forestell:2017wov}.

To enhance the self interactions of SIDM at low velocities, the case of $2 m_\chi < m_\phi \ll m_\Psi$ is of our concern. The relation $\mu \ll \lambda m_\Psi$ holds if the Yukawa couplings are similar to that of the SM Higgs boson, and the $\phi^3$-term will be negligible in SIDM annihilations. In the period of SIDM freeze-out, the main annihilation mode of SIDM is the $p-$wave process $\Psi \bar{\Psi} \to \phi \phi$, and the annihilation cross section is approximately
\begin{eqnarray}
\sigma_\mathrm{ann} v_\mathrm{r}  \approx  \frac{1}{2}  \frac{\lambda^4 (s - 4 m_\Psi^2)}{48 \pi (s - 2 m_\Psi^2) s^2}  (s + 32 m_\Psi^2)  ~,  \label{ann-phi}
\end{eqnarray}
where $v_\mathrm{r}$ is the relative velocity between the two SIDM particles. The factor $\frac{1}{2}$ is for the $\Psi \bar{\Psi}$ pair required in SIDM annihilations. $s$ is the total invariant mass squared, with $s$ = 4$m_\Psi^2 + m_\Psi^2 v_\mathrm{r}^2 + \mathcal{O} (v_\mathrm{r}^4)$. In Eq. (\ref{ann-phi}), the terms of $\mathcal{O} (v_\mathrm{r}^4)$ are neglected. The lifetime of $\phi$ should be much smaller than a second with the constraint of the BBN. As $\phi$'s couplings to SM fermions should be very tiny to evade constraints from direct detection, and here the dark sector decay of $\phi$ predominantly decaying into $\chi \bar{\chi}$ pairs could do the job ($ m_\phi > 2 m_\chi$). In addition, the mass $m_\chi \gtrsim$ 10 MeV can be tolerated by constraints from the BBN \cite{Berlin:2018sjs,Escudero:2018mvt}, and here $m_\phi \gtrsim$ 20 MeV is adopted. For fermionic $\chi$, the decay width of $\phi$ is
\begin{eqnarray}
\Gamma_\phi \simeq  \frac{\lambda_0^2 m_\phi}{8 \pi} \bigg( 1-\frac{4 m_\chi^2}{m_\phi^2} \bigg)^{3/2}  ~ .
\end{eqnarray}
Hence a very tiny mixing $\theta_\mathrm{mix}$ between $\phi$ and SM Higgs boson is compatible with the BBN constraint, and SIDM could evade the present DM direct detection hunts.

\section{Self interactions of SIDM}

\begin{figure}[htbp!]
\includegraphics[width=0.38\textwidth]{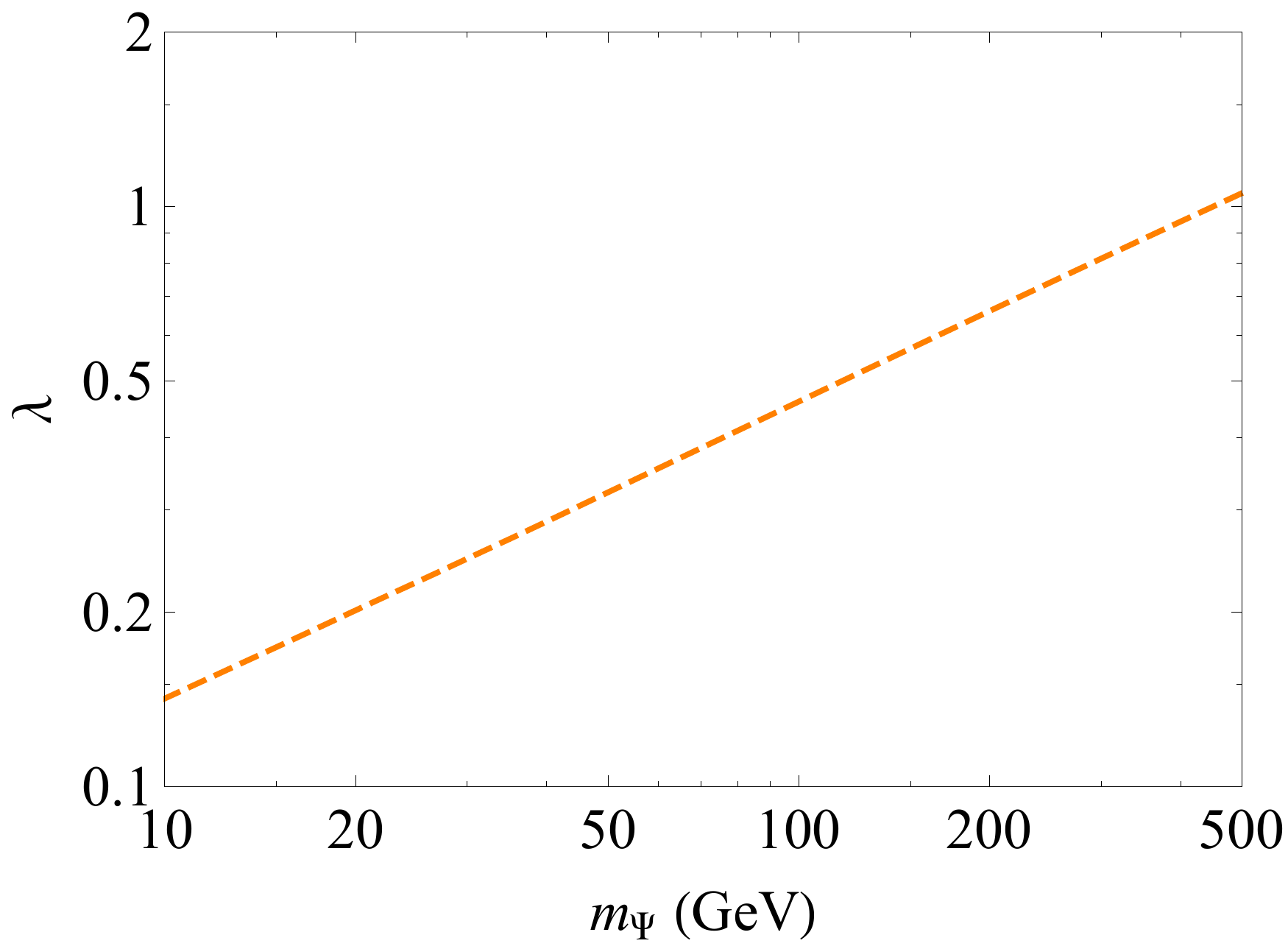} \vspace*{-1ex}
\caption{The effective coupling $\lambda$ as a function of SIDM mass $m_\Psi$, with $m_\Psi$ in a range of 10$-$500 GeV. Here the relic fraction of SIDM $f_{\mathrm{SIDM}} \simeq$ 99.6\% is taken.}\label{coupling-dm}
\end{figure}

Here we first estimate couplings set by the relic abundance of DM. The total relic abundance of DM is $\Omega_D h^2 = 0.120 \pm 0.001$ \cite{Aghanim:2018eyx}, and there are two components of DM in this paper, the main component of SIDM $\Psi$ and a small fraction of millicharged DM $\chi$. To explain the 21-cm anomaly, MeV millicharged DM with a relic fraction about 0.4\% could do the job. Thus, the relic fraction of SIDM $f_{\mathrm{SIDM}} \simeq$ 99.6\% is adopted. Taking the millicharged DM in Ref. \cite{Jia:2019yhr} as an example, the effective degree of freedom from the new sector is about 7.5 (fermionic millicharged DM, dark photon and $\phi$) at the SIDM freeze-out temperature $T_f$. Considering the relic fraction of SIDM and the effective degree of freedom \cite{Drees:2015exa} from SM + the new sector, the effective coupling $\lambda$ can be derived for a given SIDM mass $m_\Psi$, as shown in Fig. \ref{coupling-dm}. Additionally, considering the perturbative limit, $\alpha_\lambda$ ($\alpha_\lambda$ = $\lambda^2 / 4 \pi$) should be very small compared with 1.

For the case of $m_\phi \ll m_{\Psi}$, the $p-$wave annihilation $\Psi \bar{\Psi} \to \phi \phi$ with $\phi$ decaying into $\bar{\chi} \chi$ could be enhanced or suppressed at low velocities with the Sommerfeld effect considered \cite{Sommerfeld:1931,Chen:2013bi}, which is related to the mediator's mass. Note a parameter
\begin{eqnarray}
\varepsilon_\phi  \equiv  \frac{m_\phi}{\alpha_\lambda m_{\Psi}} ~.
\end{eqnarray}
In the region of $\varepsilon_\phi \lesssim 10^{-3}$, the annihilation cross section scales as $1/v_\mathrm{r}$,  and in the region of $10^{-3} \lesssim \varepsilon_\phi \lesssim 10^{-1}$, the annihilation cross section has resonant behavior \cite{Chen:2013bi}. In the region of $ \varepsilon_\phi \gtrsim 10^{-1}$, the annihilation cross section scales as $v_\mathrm{r}^2$. In addition, to explain the anomalous 21-cm absorption at the cosmic dawn, the millicharged DM $\bar{\chi} \chi$ required should be colder than the neutral hydrogen. The energetic millicharged DM from low-velocity SIDM annihilations should be as small as possible, and therefore the case of $ \varepsilon_\phi \gtrsim 10^{-1}$ is of our concern. In this case, the energetic millicharged DM $\bar{\chi} \chi$ injected from low-velocity SIDM annihilations are deeply suppressed by $v_\mathrm{r}^2$, and a bound of $\phi$'s mass is $m_\phi \gtrsim  0.1 \alpha_\lambda m_{\Psi} $.

\begin{figure*}[htbp!]
\includegraphics[width=0.32\textwidth]{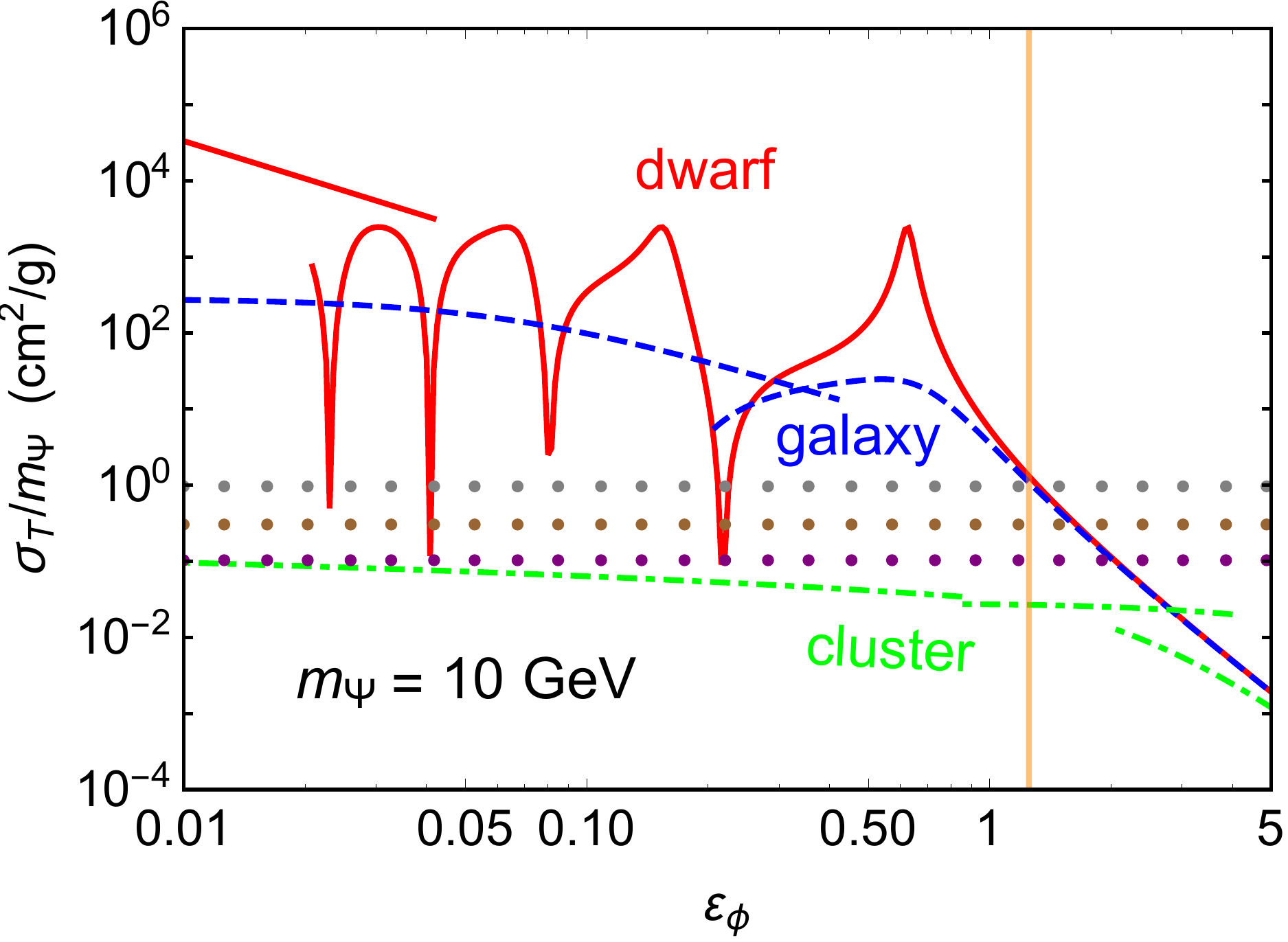} \vspace*{-1ex}
\includegraphics[width=0.32\textwidth]{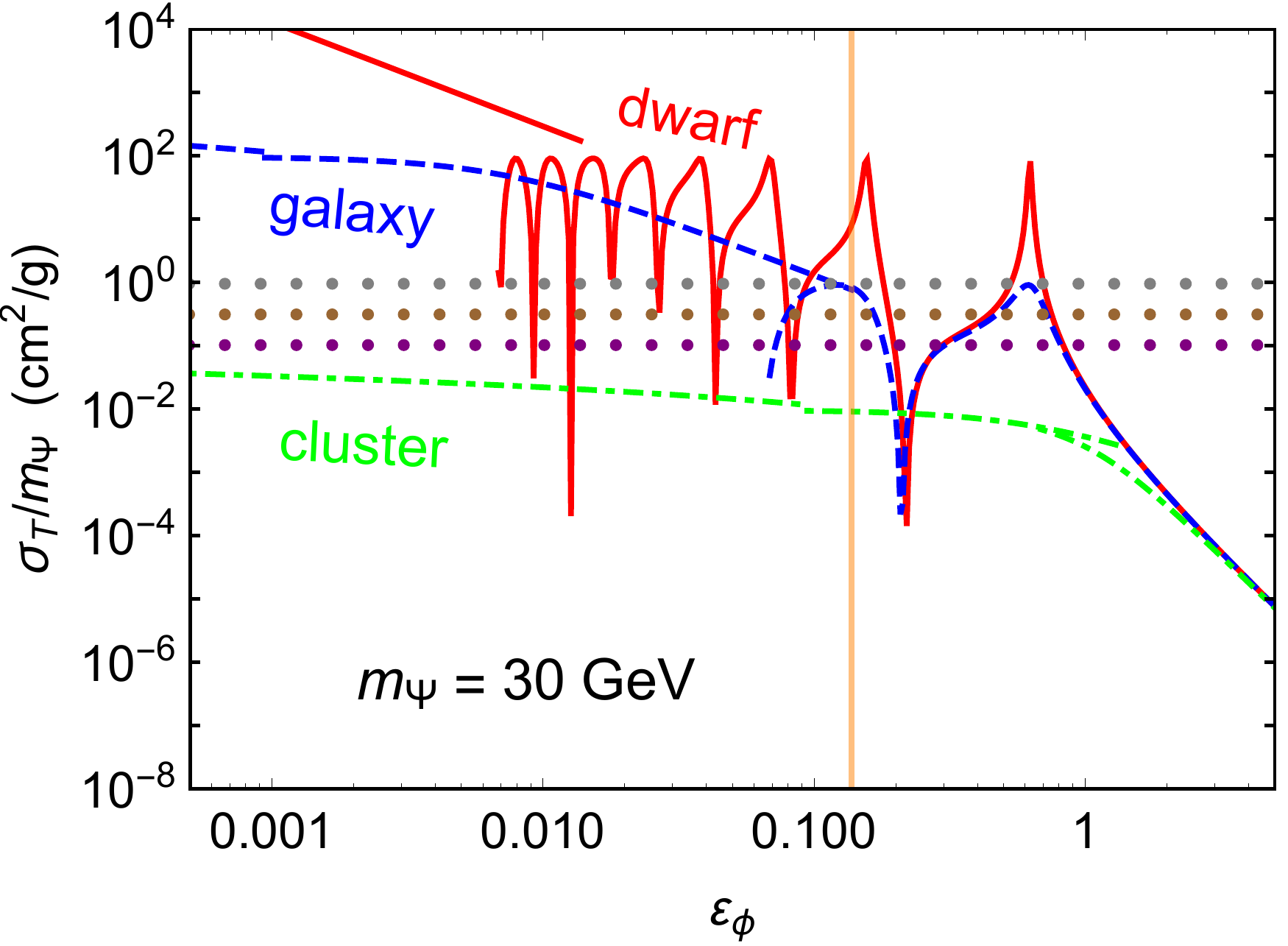} \vspace*{-1ex}
\includegraphics[width=0.32\textwidth]{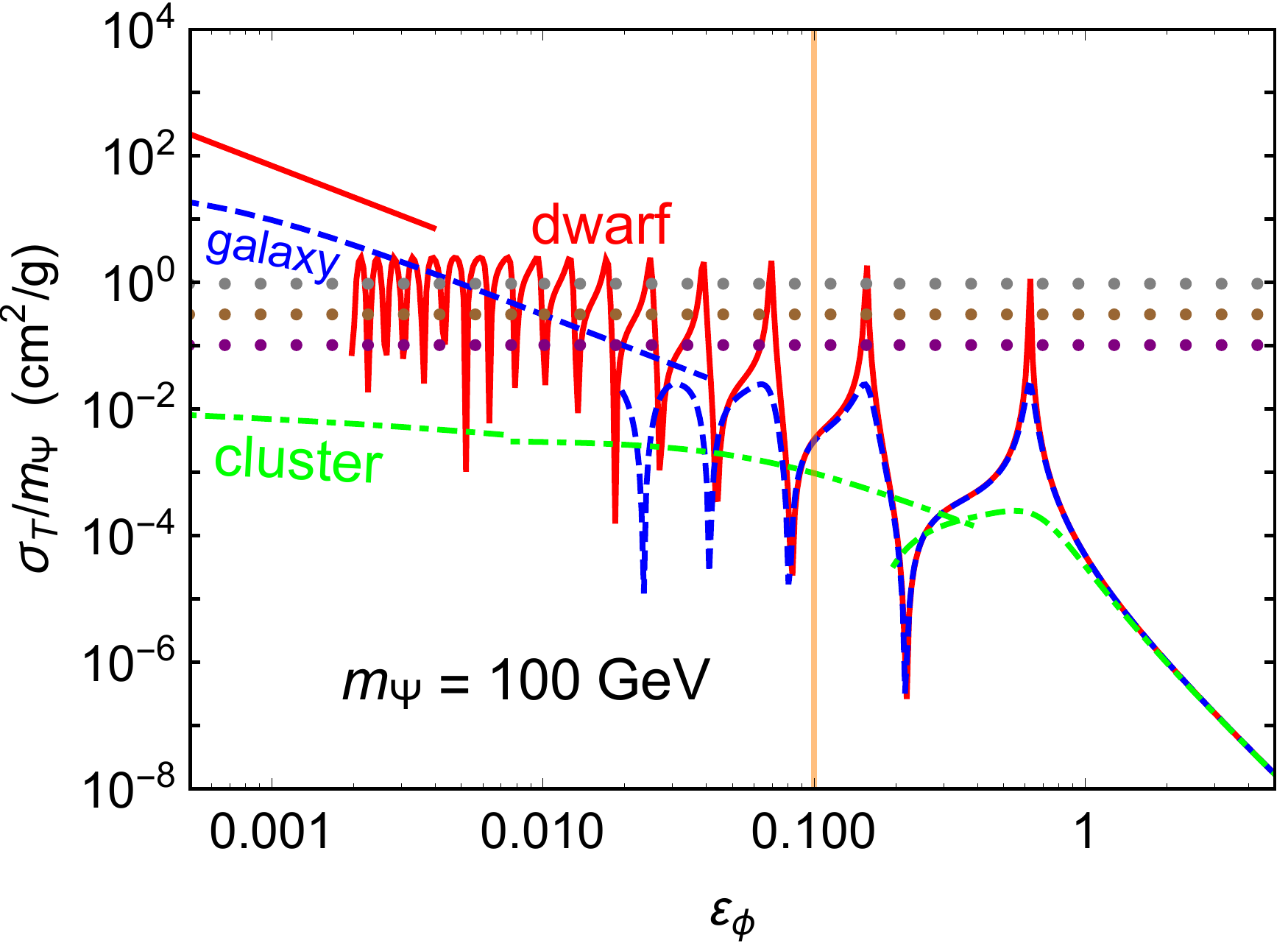} \vspace*{-1ex}
\caption{The self-scattering cross section per unit SIDM mass $\sigma_T/m_\Psi$ as a function of the parameter $\varepsilon_\phi$ at dwarf, galaxy, and cluster scales with SIDM masses $m_\Psi =$ 10, 30, 100 GeV. The solid, dashed, and dot-dashed curves are corresponding to dwarf, galaxy, and cluster scales with the typical relative velocities $v_\mathrm{r} =$ 20 km/s, 200 km/s, and 2000 km/s, respectively. For $1 \leq m_\Psi v_\mathrm{r} / m_\phi \leq 2$, both the classical result $\sigma_T^\mathrm{clas}$ and the resonant result $\sigma_T^\mathrm{Hulth\acute{e}n}$ are depicted. The dotted lines from top to bottom are for cases of $\sigma_T /m_\Psi =$ 1, 0.3, and 0.1 cm$^2$/g, respectively. The vertical lines are corresponding to a lower bound of $m_\phi$ with $m_\phi =  max \big[$20 MeV, $ 0.1 \alpha_\lambda m_{\Psi} \big] $ adopted.}\label{self-cs}
\end{figure*}

Now we turn to the self-interaction of SIDM in the non-relativistic case. The transfer cross section $\sigma_T$ in SIDM self scattering is
\begin{eqnarray}
\sigma_T = \int d \Omega (1- \cos \theta) \frac{d \sigma}{d \Omega}  ~ ,
\end{eqnarray}
and $\frac{d \sigma}{d \Omega}$ is the differential self-scattering cross section of a SIDM pair. In the Born regime ($\alpha_\lambda  m_\Psi / m_\phi \ll 1$), the cross section can be computed perturbatively, which is approximately constant for different relative velocities. To obtain an enhanced self interaction of SIDM at low velocities, the nonperturbative regime ($\alpha_\lambda  m_\Psi / m_\phi \gtrsim 1$) is considered here. Within the nonperturbative regime, for $m_\Psi v_\mathrm{r} / m_\phi \gg 1$, the result can be obtained in the classical limit, i.e., the cross section \cite{Feng:2009hw,Tulin:2013teo}
\begin{eqnarray}
\sigma_T^\mathrm{clas} \simeq \Bigg \{ \begin{array}{cc}
 \frac{4 \pi}{m_\phi^2} \beta^2 \ln (1 + \beta^{-1})            \hfill &  \beta \lesssim 0.1 \,, \\
 \frac{8 \pi}{m_\phi^2} \frac{\beta^2}{1 + 1.5 \beta^{1.65}}    \hfill &  0.1 \lesssim \beta \lesssim 10^3 \,, \\
 \frac{\pi}{m_\phi^2} (\ln \beta +1 -\frac{1}{2 \ln \beta} )^2  \hfill &  \beta \gtrsim 10^3 \,,
\end{array}
\end{eqnarray}
with $\beta \equiv 2 \alpha_\lambda m_\phi /  m_\Psi v_\mathrm{r}^2$. For $m_\Psi v_\mathrm{r} / m_\phi \lesssim 1$, an analytic result for the resonant s-wave scattering with Hulth$\acute{e}$n potential is \cite{Tulin:2013teo}
\begin{eqnarray}
\sigma_T^\mathrm{Hulth\acute{e}n} =  \frac{16 \pi}{m_\Psi^2 v_\mathrm{r}^2} \sin^2 \delta_0  ~,
\end{eqnarray}
where the phase shift $\delta_0$ is given in terms of the $\Gamma$ function, with
\begin{eqnarray}
\delta_0 = \mathrm{arg} \bigg(  \frac{i \Gamma (\lambda_+ + \lambda_- -2)}{\Gamma(\lambda_+) \Gamma(\lambda_-)}      \bigg)  ~,
\end{eqnarray}
and
\begin{eqnarray}
\lambda_\pm \equiv 1 + \frac{i m_\Psi v_\mathrm{r}}{2 \kappa m_\phi} \pm \sqrt{\frac{\alpha_\lambda m_\Psi}{\kappa m_\phi}  - \frac{ m_\Psi^2 v_\mathrm{r}^2 }{4 \kappa^2 m_\phi^2}  } ~.
\end{eqnarray}
Here the parameter $\kappa$ is $\kappa \approx$ 1.6. In nonperturbative regime, the self-interaction between SIDM particles could be enhanced at low velocities, which may resolve the small-scale problems and evade constraints from clusters. The corresponding parameter spaces will be derived in the following.

For velocity-dependent self interactions of SIDM, the typical relative velocities $v_\mathrm{r}$ in the dwarf, galaxy, and cluster scales are 20 km/s, 200 km/s and 2000 km/s, respectively. In the nonperturbative regime of $\varepsilon_\phi \lesssim 1$, the self-scattering cross section $\sigma_T$ of SIDM can be described by $\sigma_T^\mathrm{clas}$, $\sigma_T^\mathrm{Hulth\acute{e}n}$ for given relative velocities. For given SIDM masses ($m_\Psi =$ 10, 30, 100 GeV), the typical self-interactions at dwarf, galaxy, and cluster scales are shown in Fig. \ref{self-cs}. Considering $\sigma_T/m_\Psi \gtrsim$ 1 cm$^2$/g at dwarf and galaxy scales and $\sigma_T/m_\Psi \lesssim$ 0.1$-$0.3 cm$^2$/g at cluster scale, there are parameter spaces to resolve the small-scale problems and meanwhile be compatible from dwarf to cluster scales, with 10 $\lesssim m_\Psi \lesssim$ 40 GeV.

\begin{figure*}[htbp!]
\includegraphics[width=0.32\textwidth]{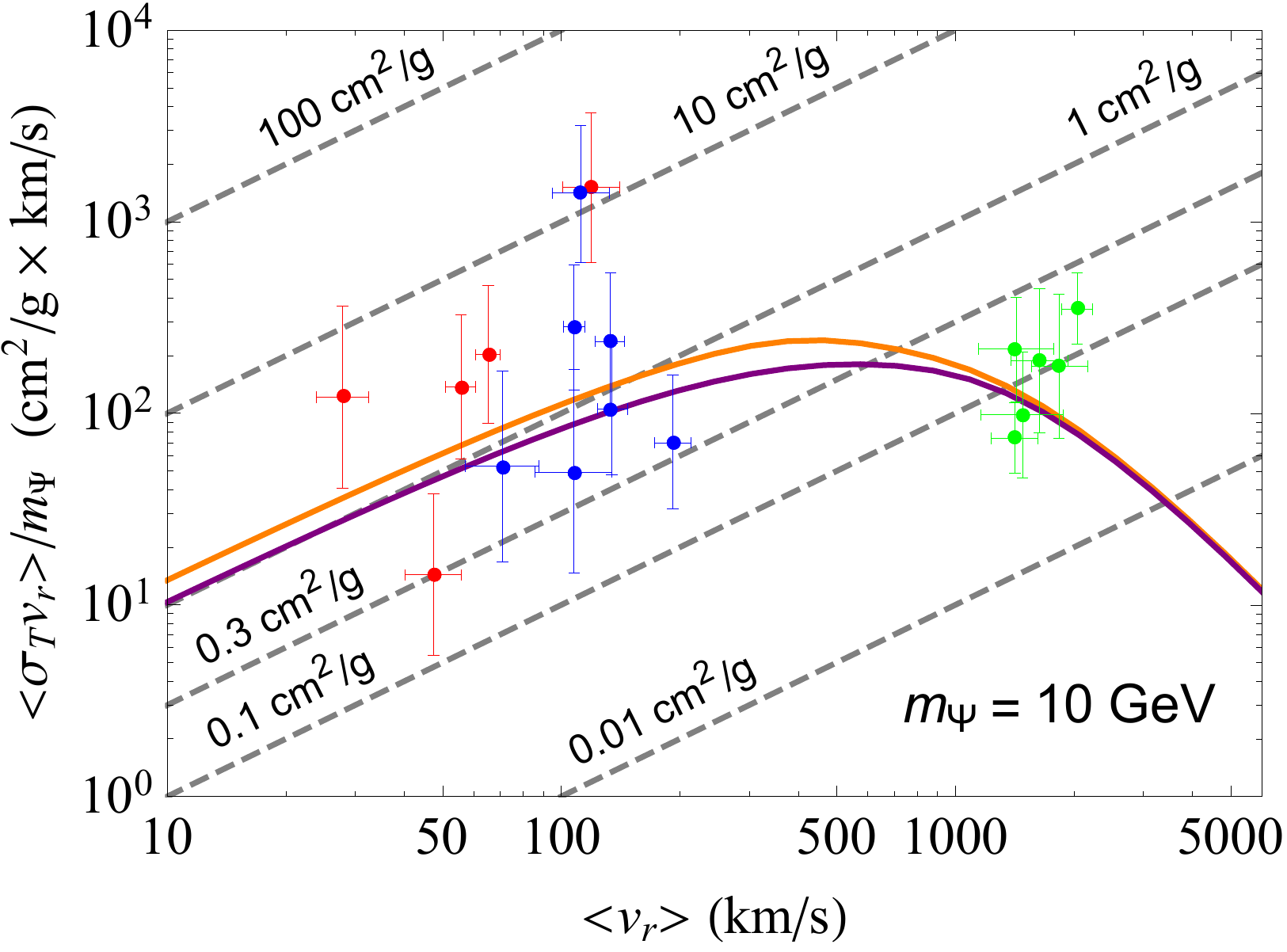} \vspace*{-1ex}
\includegraphics[width=0.32\textwidth]{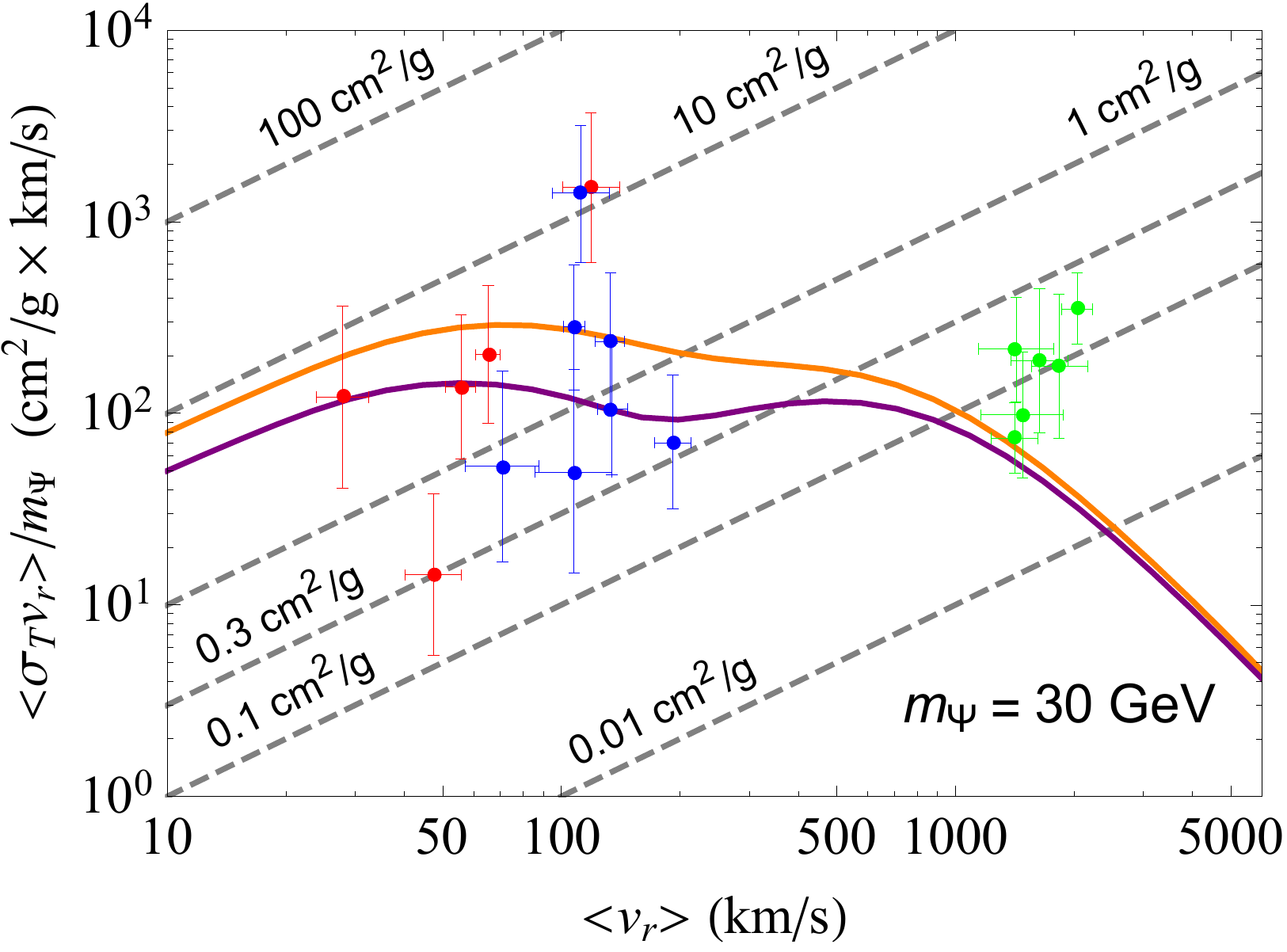} \vspace*{-1ex}
\includegraphics[width=0.32\textwidth]{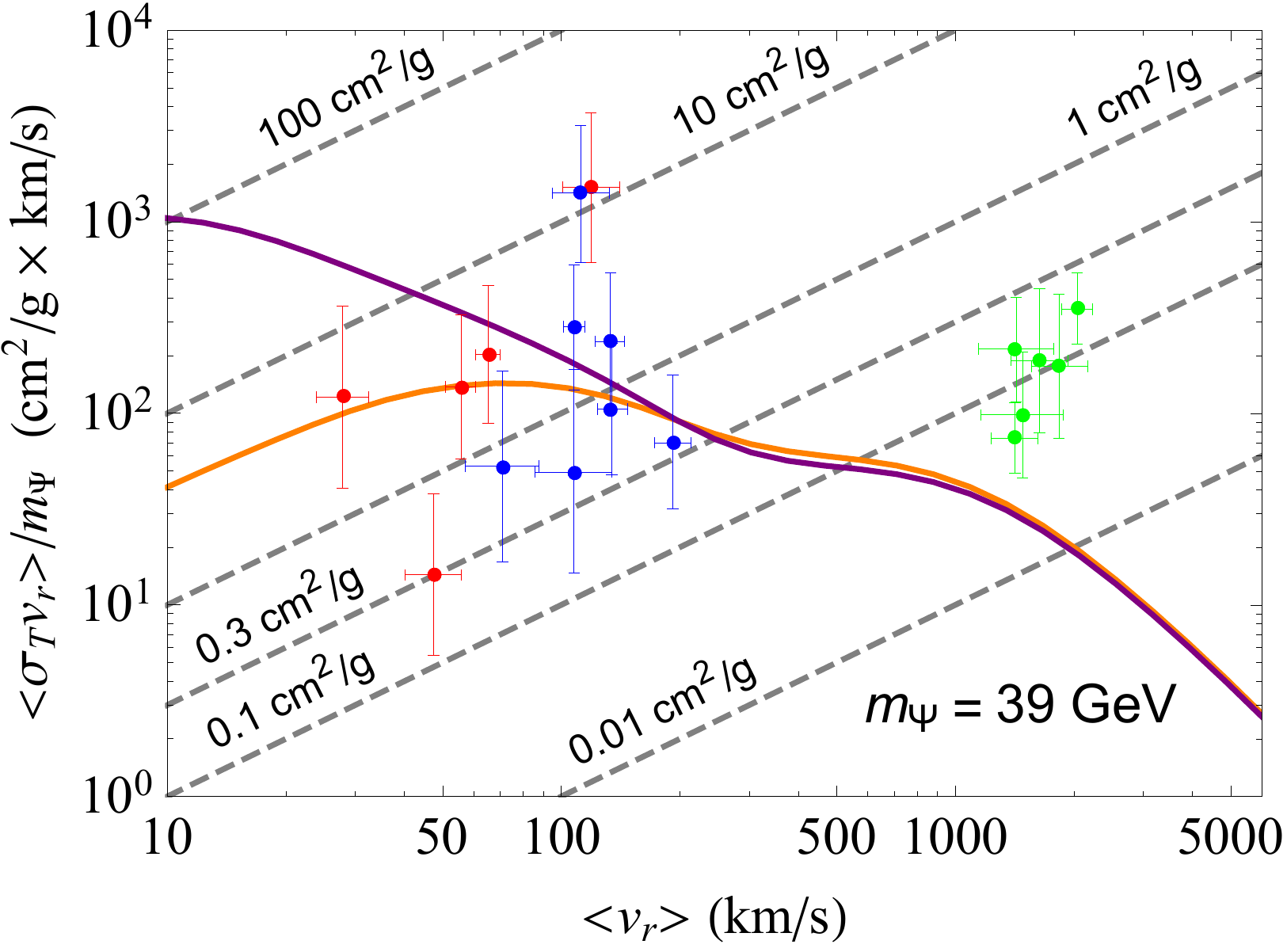} \vspace*{-1ex}
\caption{The velocity-weighted self-scattering cross section per unit SIDM mass $\langle \sigma_T v_r \rangle/m_\Psi$ as a function of the mean relative velocity $\langle v_r\rangle$ at given SIDM masses $m_\Psi =$ 10, 30 and 39 GeV. The solid curves for each SIDM mass are values of $\langle \sigma_T v_r \rangle/m_\Psi$, with the orange one, purple one corresponding to the lower bound, upper limit of $m_\phi$'s range being taken, respectively. The averaged cross section per unit SIDM mass is taken as $\langle \sigma_T v_r \rangle/( \langle v_r \rangle m_\Psi ) \gtrsim$ 1 cm$^2$/g at dwarf and galaxy scales and $\langle \sigma_T v_r \rangle/( \langle v_r \rangle m_\Psi ) \lesssim$ 0.1$-$0.3 cm$^2$/g at cluster scale. Additionally, the upper limit of $\langle \sigma v_r \rangle/( \langle v_r \rangle m_\Psi )$ is also constrained by the lower bound of $\phi$'s mass $m_\phi \gtrsim $ 20 MeV. The points are the inferred values of $\langle \sigma_T v_r \rangle/m_\Psi$ from dwarfs (red), low surface brightness galaxies (blue) and clusters (green) \cite{Kaplinghat:2015aga}. The dashed lines from top left to bottom right are for the case of constant $\sigma_T/m_\Psi$ with $\sigma_T/m_\Psi =$ 100, 10, 1, 0.3, 0.1 and 0.01 cm$^2$/g, respectively.}\label{sig-average}
\end{figure*}

\begin{figure}[htbp!]
\includegraphics[width=0.38\textwidth]{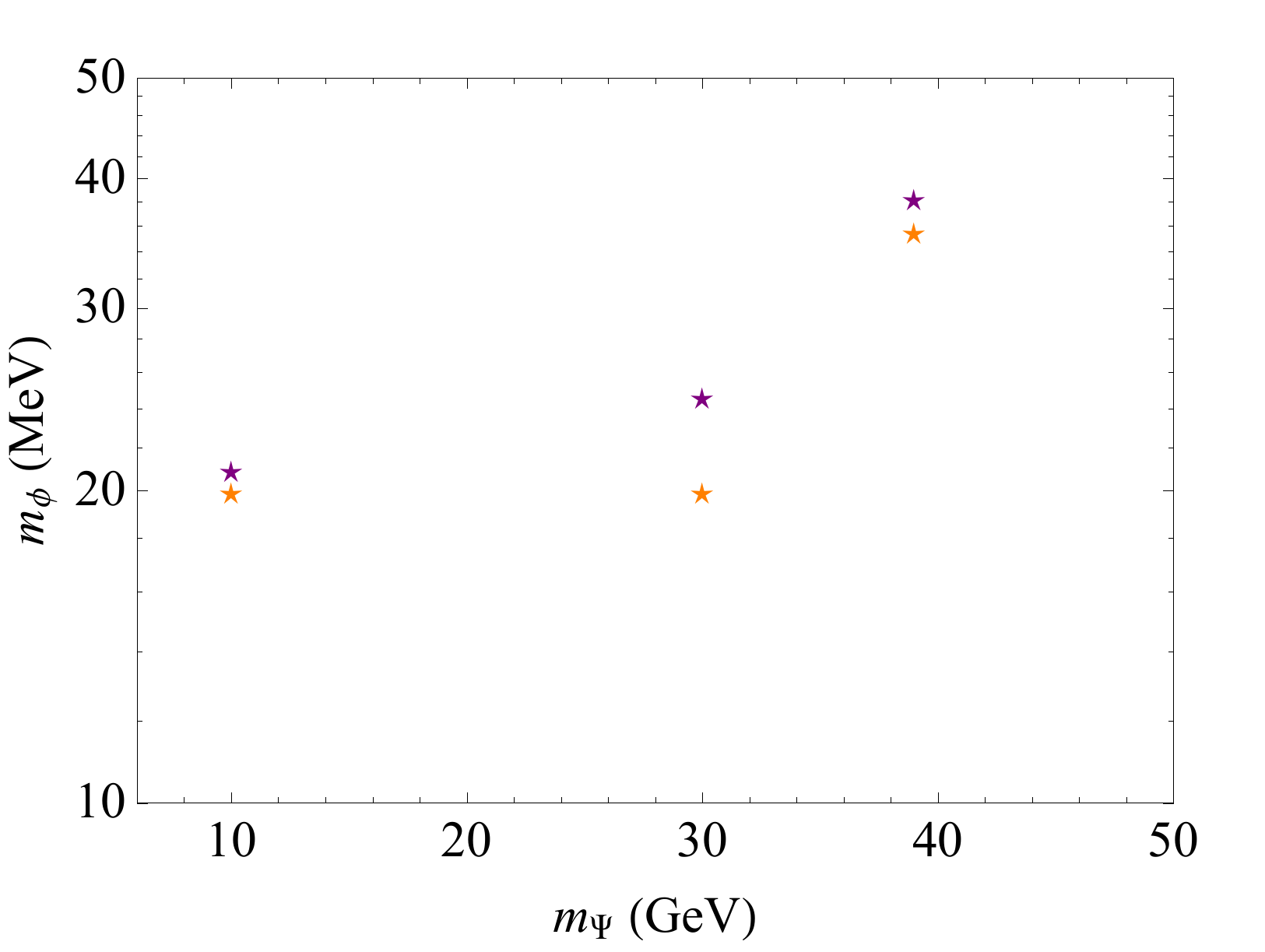} \vspace*{-1ex}
\caption{The values of $m_\phi$ required for given SIDM masses of $m_\Psi =$ 10, 30 and 39 GeV (Fig. \ref{sig-average}). The stars are the lower bounds (orange) and upper limits (purple) of $m_\phi$.}\label{eps-mph}
\end{figure}

In the above self-interactions of SIDM, the monochromatic typical relative velocities $v_\mathrm{r}$ are adopted in the dwarf, galaxy, and cluster scales. Actually, the distribution of SIDM velocities needs to be taken into account, and this will give a mild modification. In the inner regions of dwarf galaxies, galaxies, and clusters, the inner profile is related to the velocity-averaged self-scattering cross section per unit of SIDM mass $\langle \sigma_T v_\mathrm{r}\rangle /m_\Psi$ \cite{Kaplinghat:2015aga}, where
\begin{eqnarray}
\langle \sigma_T v_\mathrm{r} \rangle = \int_0^{v_\mathrm{r}^{\mathrm{max}}}  f(v_\mathrm{r}, v_0) \sigma_T v_\mathrm{r}  d v_\mathrm{r}  ~,
\end{eqnarray}
and a Maxwell-Boltzmann velocity distribution is assumed, with
\begin{eqnarray}
f(v_\mathrm{r}, v_0) =  \frac{4 v_\mathrm{r}^2 e^{- v_\mathrm{r}^2/ v_0^2}}{\sqrt{\pi} v_0^3}  ~.
\end{eqnarray}
The escape velocity can be taken as $v_\mathrm{r}^{\mathrm{max}}$, and $v_0$ is a parameter related to the typical velocities in the DM halo. In the inner regions of halos, $v_\mathrm{r}^{\mathrm{max}}$ is much larger than $v_0$, and the averaged relative velocity $\langle v_\mathrm{r} \rangle$ is $\langle v_\mathrm{r} \rangle \simeq$ 2$v_0/\sqrt{\pi}$. Here we take the averaged self-interaction cross section as $\langle \sigma_T v_r \rangle/ \langle v_r \rangle$, and adopt the constraints of $\langle \sigma_T v_r \rangle/( \langle v_r \rangle m_\Psi ) \gtrsim$ 1 cm$^2$/g at dwarf and galaxy scales and $\langle \sigma_T v_r \rangle/( \langle v_r \rangle m_\Psi ) \lesssim$ 0.1$-$0.3 cm$^2$/g at cluster scale. Considering the velocity distributions, the self-interactions of SIDM at dwarf, galaxy, and cluster scales are shown in Fig. \ref{sig-average}, with $m_\Psi =$ 10, 30 and 39 GeV, and the corresponding ranges of $m_\phi$ are shown in Fig. \ref{eps-mph}. It can be seen that, there are parameter spaces to resolve the small scale problems and meanwhile be compatible with constraints from clusters.

\section{Direct detection of SIDM}

Now we turn to the direct detection of SIDM. In WIMP-type DM direct detections, the momentum transfer $|q|$ in the WIMP-target nucleus elastic scattering is generally assumed to be much smaller than the mediator mass $m_{\mathrm{med}}$, and thus the WIMP-nucleus elastic scattering cross section could be derived in the limit of zero momentum transfer $|q^2| \rightarrow$ 0. The $q$-dependent squared matrix element for WIMP-nucleus spin-independent (SI) elastic scattering $|\mathcal{M}_{\Psi N} (q)|^2$ can be written as
\begin{eqnarray}
|\mathcal{M}_{\Psi N} (q)|^2 &=& |\mathcal{M}_{\Psi N} (q)|^2 |_{q^2=0}  \frac{m_{\mathrm{med}}^4}{(|q^2| + m_{\mathrm{med}}^2)^2}     \nonumber \\
&& \times   |F_N^{\mathrm{SI}} (q)|^2     ~,
\end{eqnarray}
where $F_N^{\mathrm{SI}} (q)$ is the nuclear form factor. For a small momentum transfer with $1/|q|$ larger than the nuclear radius, the nuclear form factor is $|F_N^{\mathrm{SI}} (q)|^2 \rightarrow 1$. Note
\begin{eqnarray}
 F_{\mathrm{med}}(q^2) =  \frac{m_\mathrm{med}^4}{(|q^2| + m_\mathrm{med}^2)^2}  .
\end{eqnarray}
In the limit of $|q^2|/m_\mathrm{med}^2 \rightarrow$ 0, one has $F_{\mathrm{med}}(q^2) \simeq$ 1.\footnote{In the case of $F_{\mathrm{med}}(q^2) \approx $ 1, the $q$-dependent nuclear form factor $F_N^{\mathrm{SI}} (q)$ needs to be considered for heavy nuclei.} Thus, the WIMP-nucleus scattering is a contact interaction, and a constant WIMP-nucleus scattering cross section can be extracted from the recoil rate \cite{Lewin:1995rx}, without consideration of the mediator's mass. For the scalar mediator $\phi$ of concern, $m_\phi/ m_\Psi$ is $\sim 10^{-3}$, and the velocity of the incoming SIDM $v_\mathrm{in}$ relative to the Earth detector is $v_\mathrm{in}/c  \sim 10^{-3}$. Therefore, the zero momentum transfer limit fails in direct detections.

In GeV SIDM-target nucleus elastic scattering, the target nucleus can be considered to be at rest initially, and the momentum transfer is $q \rightarrow (0, \vec{q})$. The nucleus recoil energy $E_R$ is
\begin{eqnarray}
E_R = \frac{\mu_{\Psi N}^2 v_\mathrm{in}^2}{m_N} (1-\cos \theta_{\mathrm{cm}}) = \frac{|\vec{q}|^2}{2 m_N} ~,
\end{eqnarray}
where $m_N$ is the target nucleus mass, $\mu_{\Psi N}^{}$ is the reduced mass of the SIDM-nucleus system, and $\theta_{\mathrm{cm}}$ is the polar angle in the center-of-momentum frame in the SIDM-nucleus scattering. For a given recoil energy $E_R$, the minimum incoming velocity of SIDM is $v_\mathrm{in}^\mathrm{min} = \sqrt{m_N E_R / 2 \mu_{\Psi N}^2}$. The available maximum value $|\vec{q}|^2_{\mathrm{max}}$ are related to the maximum velocity squared $(v_\mathrm{in}^2)_{\mathrm{max}}$ and the maximum nuclear recoil energy $E_R^\mathrm{max}$ in DM detections. For SIDM with the escape velocity $v_\mathrm{esc}$, the SIDM incoming velocity squared is $v_\mathrm{in}^2 \approx v_\mathrm{esc}^2 + v_{\oplus}^2 - 2 v_\mathrm{esc} v_{\oplus} \cos \theta$, where $v_{\oplus}$ is the Earth's velocity relative to the galactic center (the influence of the Earth annual modulation is not taken into account). The values of $v_\mathrm{esc} =$ 544 km/s and $v_{\oplus} =$ 232 km/s are adopted. For DM direct detection experiments, the results from XENON1T \cite{Aprile:2018dbl}, LUX \cite{Akerib:2016vxi}, and PandaX-II \cite{Cui:2017nnn} set strong limits on WIMP type DM with masses $\gtrsim$ 10 GeV. Here the nucleus recoil energy region of interest in the XENON1T experiment \cite{Aprile:2018dbl}, i.e. [4.9, 40.9] keV$_{\mathrm{nr}}$, is employed to set the range of $|\vec{q}|^2$ in calculations.

In the SIDM-target nucleus SI elastic scattering, the differential cross section can be evaluated as
\begin{eqnarray}
\frac{d \sigma_N^{\mathrm{SI}} (q)}{d E_R}  = \frac{m_N}{2 \mu_{\Psi N}^2 v_\mathrm{in}^2} \sigma_N^{\mathrm{SI}}(q)|_{q^2=0}  F_{\mathrm{med}}(q^2)  |F_N^{\mathrm{SI}} (q)|^2  \, ,   \nonumber \\   \label{diff-cs}
\end{eqnarray}
with $|q| = \sqrt{2 m_N  E_R}$. The SIDM-nucleus scattering cross section at $q^2 \rightarrow$ 0 is
\begin{eqnarray}
\sigma_N^{\mathrm{SI}}(q)|_{q^2=0} &=& \sigma_p^{\mathrm{SI}}|_{q^2=0} \frac{\mu_{\Psi N}^2}{\mu_{\Psi p}^2}   \nonumber \\
&& \times [Z + \frac{f_n}{f_p} (A-Z)]^2 ,
\end{eqnarray}
where $\sigma_p|_{q^2=0}$ is the SIDM-proton scattering cross section in the limit of $q^2=0$, $\mu_{\Psi p}$ is the SIDM-proton reduced mass. $Z$ is the number of protons, $A$ is the mass number of the nucleus, and ${f_n}$ and ${f_p}$ describe the SIDM-neutron and SIDM-proton couplings respectively. For $\phi$-mediated scattering, one has ${f_n} = {f_p}$, and the SIDM-nucleon elastic scattering cross section can be defined as
\begin{eqnarray}
\sigma_n^{\mathrm{SI}} \equiv \sigma_p^{\mathrm{SI}}|_{q^2=0}   ~ .
\end{eqnarray}
Now, Eq. (\ref{diff-cs}) can be rewritten as
\begin{eqnarray}
\frac{d \sigma_N^{\mathrm{SI}} (q)}{d E_R}  = \frac{m_N}{2 \mu_{\Psi p}^2 v_\mathrm{in}^2} \sigma_n^{\mathrm{SI}}  F_{\mathrm{med}}(q^2)  A^2  |F_N^{\mathrm{SI}} (q)|^2  \, .
\label{diff-csnr}
\end{eqnarray}

Here a reference value of $F_{\mathrm{med}}(q^2)$ is introduced in direct detections, i.e., a reference factor $\overline{F}_{\mathrm{med}}$. For all target nuclei in one species, the factor $\overline{F}_{\mathrm{med}}$ is
\begin{eqnarray}
\overline{F}_\mathrm{med}  =
\frac{ \int_{E_R^\mathrm{thr}}^{E_R^\mathrm{high}}  d E_R ~ \epsilon(E_R) \frac{d R }{d E_R} }
{\int_{E_R^\mathrm{thr}}^{E_R^\mathrm{high}} d E_R ~ \epsilon(E_R) \frac{d R }{d E_R}|_{F_{\mathrm{med}}(q^2) = 1}}   ~  ,
\label{f-med}
\end{eqnarray}
where $\epsilon(E_R)$ is the detection efficiency for a given recoil energy $E_R$, and $\frac{d R }{d E_R}$ is the differential recoil rate (see the Appendix \ref{kq} for the details). For target nuclei in the same species, we have
\begin{eqnarray}
\overline{F}_\mathrm{med}  =
\frac{ \int_{E_R^\mathrm{thr}}^{E_R^\mathrm{high}}  d E_R ~ \epsilon(E_R) F_\mathrm{med}(q^2) |F_N^{\mathrm{SI}} (q)|^2  \eta ( v_\mathrm{in}^\mathrm{min}) }
{\int_{E_R^\mathrm{thr}}^{E_R^\mathrm{high}} d E_R ~ \epsilon(E_R)  |F_N^\mathrm{SI} (q)|^2  \eta ( v_\mathrm{in}^\mathrm{min})}   ~  .   \nonumber   \\
\end{eqnarray}
The WIMP search results $\sigma_\mathrm{n}^\mathrm{SI}$ (WIMP) for WIMP-nucleus elastic scatterings with contact interactions, can be converted into the SIDM search results $\sigma_\mathrm{n}^\mathrm{SI}$ (SIDM) via the relation
\begin{eqnarray}
\sigma_\mathrm{n}^\mathrm{SI} (\mathrm{WIMP}) \simeq  f_\mathrm{SIDM}  \sigma_\mathrm{n}^\mathrm{SI} (\mathrm{SIDM}) \times \overline{F}_\mathrm{med}      ~  . \label{WIMP-SIDM}
\end{eqnarray}

\begin{figure}[htbp!]
\includegraphics[width=0.38\textwidth]{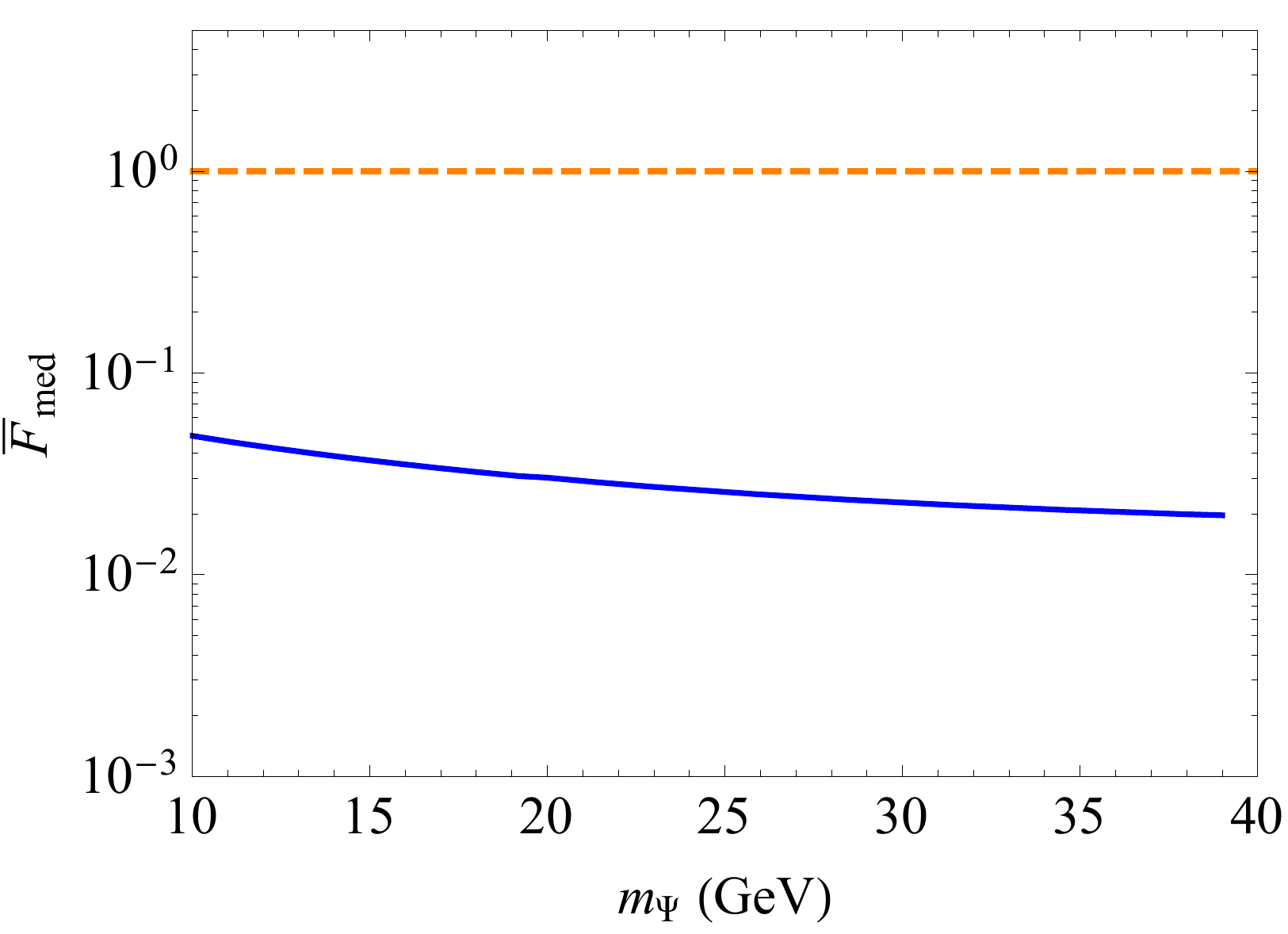} \vspace*{-1ex}
\caption{The reference factor $\bar{F}_\mathrm{med}$ as a function of SIDM mass $m_\Psi$ in SIDM-target nucleus ($^{131}$Xe) SI elastic scattering. The solid curve is the result of $\bar{F}_\mathrm{med}$, with SIDM mass in a range of 10$-$39 GeV and $m_\phi =$ 20 MeV. Here the range of the recoil energy $E_R$ and the detection efficiency $\epsilon(E_R)$ in XENON1T (2018) experiment \cite{Aprile:2018dbl} are adopted as inputs. For comparison, the dashed line is for the case $\bar{F}_\mathrm{med} =$ 1. }\label{f-mediator}
\end{figure}

\begin{figure}[htbp!]
\includegraphics[width=0.38\textwidth]{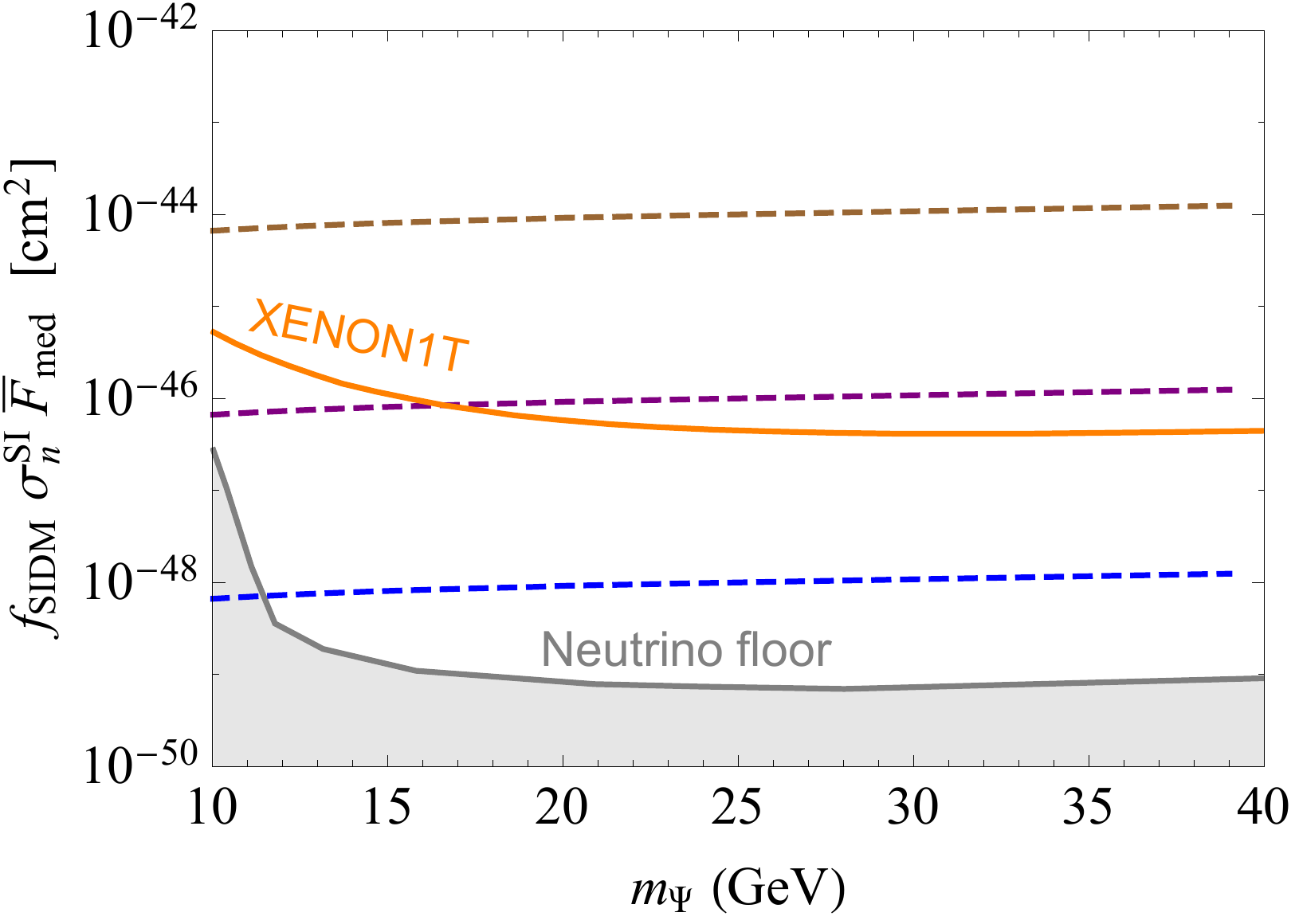} \vspace*{-1ex}
\caption{The SIDM-nucleon SI scattering cross section in the form of $f_\mathrm{SIDM}  \sigma_\mathrm{n}^\mathrm{SI}  \overline{F}_\mathrm{med}$ as a function of SIDM mass, with $m_\phi =$ 20 MeV. The dashed curves from top to bottom are the scattering cross section $f_\mathrm{SIDM}  \sigma_\mathrm{n}^\mathrm{SI}  \overline{F}_\mathrm{med}$ for the case of $\theta_\mathrm{mix} = 10^{-7}, 10^{-8}, 10^{-9}$ respectively. The upper, lower solid curves are the upper limit from XENON1T \cite{Aprile:2018dbl}, the detection bound of the neutrino floor \cite{Billard:2013qya} respectively.}\label{SIDM-direct-d}
\end{figure}

For the WIMP search result of XENON1T (2018) \cite{Aprile:2018dbl}, the detection efficiency $\epsilon(E_R)$ in the nuclear recoil energy region of interest is released (Fig. 1 in Ref. \cite{Aprile:2018dbl}). To estimate the SIDM-nucleus scattering, here $m_\phi =$ 20 MeV is adopted as an input. After substituting values of the corresponding parameters, the results of $\overline{F}_\mathrm{med}$ can be derived, as shown in Fig. \ref{f-mediator}. It can be seen that, the approximation of contact interactions between WIMP-nucleus scatterings fails, and the mediator's mass needs considered in direct detections of SIDM.

Now we launch a specific WIMP detection result (XENON1T-2018 \cite{Aprile:2018dbl}) to the SIDM of concern. The cross section of SIDM-nucleon (proton, neutron) SI elastic scattering mediated by $\phi$ can be parameterized as
\begin{eqnarray}
\sigma_n^\mathrm{SI}  = \frac{\lambda^2  \theta_\mathrm{mix}^2  g_{hnn}^2 }{\pi m_\phi^4}  \mu_{\Psi p}^2    ~  ,
\end{eqnarray}
where $g_{hnn}^{}$ is the effective Higgs-nucleon coupling, with $g_{hnn}^{} \simeq 1.1 \times 10^{-3}$ \cite{Cheng:2012qr} adopted here. For SIDM-nucleus scattering with a light mediator $\phi$, though the cross section $\sigma_n^\mathrm{SI}$ cannot be directly extracted from the recoil rate in direct detections, the factor $f_\mathrm{SIDM}  \sigma_\mathrm{n}^\mathrm{SI}  \overline{F}_\mathrm{med}$ is feasible, as discussed above. Here we take $^{131}$Xe as the target nucleus of the liquid xenon detector for simplicity. Considering the constraint of WIMPs from XENON1T \cite{Aprile:2018dbl}, the result for SIDM detection is shown in Fig. \ref{SIDM-direct-d}. For SIDM with masses in a range of 10$-$39 GeV, the parameter $\theta_\mathrm{mix}$ should be $\lesssim 10^{-8}$.

In addition, for given SIDM and light mediator masses, constraints on WIMP-nucleon scattering cross section derived by different DM detection experiments cannot be directly applied to the SIDM detection in company, and this is due to the value of $\overline{F}_\mathrm{med}$ being related to some characters of the detectors, i.e., the constituent of target material, the nucleus recoil energy region of interest and corresponding detection efficiency. In this case, the scattering cross section $f_\mathrm{SIDM}  \sigma_\mathrm{n}^\mathrm{SI} (\mathrm{SIDM})$ is available for comparison between different detection experiments, with
\begin{eqnarray}
\frac{ \sigma_\mathrm{n}^\mathrm{SI} (\mathrm{WIMP})}{\overline{F}_\mathrm{med}}  \simeq  f_\mathrm{SIDM}  \sigma_\mathrm{n}^\mathrm{SI} (\mathrm{SIDM})   ~  ,
\end{eqnarray}
i.e. the WIMP-nucleon scattering cross section divided by the factor $\overline{F}_\mathrm{med}$.

\section{Conclusion and discussion}

We have investigated a scenario of two-component DM, a small fraction is MeV millicharged DM which could cause the anomalous 21-cm absorption at the cosmic dawn, and the main component is SIDM which could resolve small-scale problems. We focus on the main component of DM, i.e. the SIDM, in this paper. The velocity-dependent self interaction of SIDM mediated by a light scalar $\phi$ has been considered, which can be compatible from dwarf to cluster scales, with SIDM's mass $m_\Psi$ in a range of 10$-$39 GeV and the mediator's mass required $m_\phi \sim 20-40$ MeV. For fermionic SIDM $\Psi$, the main annihilation mode of $\Psi \bar{\Psi} \to \phi \phi$ is a $p -$wave process. As the thermal equilibrium between SIDM and the SM particles in the very early universe via the transition of SIDM $\rightleftarrows \phi \rightleftarrows$ SM particles has been excluded by the present DM direct detections, here we considered the case that SIDM was in the thermal equilibrium with millicharged DM and $\phi$ predominantly decaying into a pair of millicharged DM. Thus, SIDM could be in the thermal equilibrium with SM particles via millicharged DM, and the $\phi -$SM particle couplings could be very tiny and evade present DM direct detections.

Due to the small mediator's mass required by the velocity-dependent self interactions of SIDM, the picture of WIMP-target nucleus scattering with contact interactions fails for SIDM-target nucleus scattering with a light mediator, and thus the detection results for WIMPs cannot be directly applied to the SIDM detection. A method is explored in this paper, with which the results of $\sigma_\mathrm{n}^\mathrm{SI}$ (WIMP) in direct detection experiments can be converted into the SIDM search results $\sigma_\mathrm{n}^\mathrm{SI}$ (SIDM), i.e., for given SIDM and mediator masses, a mediator-dependent factor $\overline{F}_\mathrm{med}$ included. With this method, the XENON1T result is employed to constrain the SIDM-nucleon SI scattering. The value of $\overline{F}_\mathrm{med}$ is related to the constituent of target material, the nucleus recoil energy region of interest and corresponding detection efficiency. It is welcome to release the nucleus recoil energy $E_R$'s region of interest and the corresponding detection efficiency $\epsilon (E_R)$ in DM direct detection experiments, and thus the WIMP detection result can be employed in SIDM hunts. We look forward to the search of SIDM in GeV scale by the future DM direct detections, such as PandaX-4T \cite{Zhang:2018xdp}, XENONnT \cite{Aprile:2015uzo}, LZ \cite{Akerib:2018lyp}, DarkSide-20k \cite{Aalseth:2017fik} and DARWIN \cite{Aalbers:2016jon}, and the detections will reach the neutrino floor in the next decade(s).

\acknowledgments    \vspace*{-3ex}


This work was supported by the National Natural Science Foundation of China under Contract No. 11505144, and the Longshan academic talent research supporting program of SWUST under Contract No. 18LZX415.



\appendix*


\section{The $\overline{F}_\mathrm{med}$} \label{kq}

To evaluate the reference factor $\overline{F}_{\mathrm{med}}$, i.e., a typical value of $F_{\mathrm{med}}(q^2)$ in direct detections, we start from the recoil rate for the SIDM-target nucleus SI elastic scattering. The differential recoil rate per unit target mass and per unit time is
\begin{eqnarray}
\frac{d R }{d E_R}  &=& \frac{\rho_\mathrm{DM}^{} f_{\mathrm{SIDM}} }{m_N  m_\Psi} \int \int \int d^3 \vec{v}_\mathrm{in} ~ [ \frac{d \sigma_N^{\mathrm{SI}} (q)}{d E_R} v_\mathrm{in}  f_\mathrm{E}(\vec{v}_\mathrm{in})     \nonumber     \\
&&  \times  \Theta (v_\mathrm{in} - v_\mathrm{in}^\mathrm{min})  ]   \, ,
\label{diff-rate}
\end{eqnarray}
where $\rho_\mathrm{DM}^{}$ is the local DM density, $f_\mathrm{E}(\vec{v}_\mathrm{in})$ is the velocity distribution of SIDM relative to the Earth, and $\Theta (v_\mathrm{in} - v_\mathrm{in}^\mathrm{min})$ is the step function corresponding to the minimum incoming velocity of SIDM for a recoil energy $E_R$. Substituting Eq. (\ref{diff-csnr}) into Eq. (\ref{diff-rate}), we have
\begin{eqnarray}
\frac{d R }{d E_R}  &=& \frac{\rho_\mathrm{DM}^{} f_{\mathrm{SIDM}} }{ m_\Psi} \frac{ A^2}{2 \mu_{\Psi p}^2 }    \sigma_n^{\mathrm{SI}}  F_{\mathrm{med}}(q^2) |F_N^{\mathrm{SI}} (q)|^2  \nonumber     \\
&&  \times  \eta ( v_\mathrm{in}^\mathrm{min})       ~  ,
\label{diff-rate-II}
\end{eqnarray}
where $\eta (  v_\mathrm{in}^\mathrm{min})$ is
\begin{eqnarray}
\eta (  v_\mathrm{in}^\mathrm{min}) =  \int \int \int d^3 \vec{v}_\mathrm{in}  \frac{f_\mathrm{E}(\vec{v}_\mathrm{in})}{v_\mathrm{in}}   \Theta (v_\mathrm{in} - v_\mathrm{in}^\mathrm{min})   ~ .
\end{eqnarray}
The incoming velocity of SIDM $\vec{v}_\mathrm{in}$ is related to the SIDM's velocity $\vec{v}_\mathrm{halo}$ in the halo via $\vec{v}_\mathrm{in} = \vec{v}_\mathrm{halo} - \vec{v}_{\oplus}$ (here the orbital motion of the Earth is neglected). For SIDM in the halo, the SIDM particles are assumed to be isotropic with a Maxwell-Boltzmann distribution,
\begin{eqnarray}
f_\mathrm{halo}(\vec{v}_\mathrm{halo}) = \frac{1}{N_\mathrm{F}} \mathrm{exp} \big( - \frac{\vec{v}_\mathrm{halo}^2}{ v_c^2} \big)  ~ ,
\end{eqnarray}
where $N_\mathrm{F}$ is the normalization factor, and the value of $v_c$ is $v_c \approx$ 220 km/s. Boosting this distribution to the Earth rest frame, one has
\begin{eqnarray}
f_\mathrm{E}(\vec{v}_\mathrm{in}) = \frac{1}{N_\mathrm{F}} \mathrm{exp} \big( - \frac{( \vec{v}_\mathrm{in} + \vec{v}_{\oplus})^2}{v_c^2} \big)  ~ .
\end{eqnarray}

A usual choice of the nuclear form factor $F_N^\mathrm{SI} (q)$ is the analytical Helm form factor \cite{Helm:1956zz,Lewin:1995rx}, which can be expressed as
\begin{eqnarray}
F_N^\mathrm{SI} (q) = \frac{3}{r_N  |q|} j_1 (r_N  |q|)  e^{- |q^2| s_\mathrm{skin}^2 /2}   ~  ,
\end{eqnarray}
where $s_\mathrm{skin}$ is the nuclear skin thickness parameter, with $s_\mathrm{skin} \approx$ 0.9 fm. $j_1 (x)$ ($x = r_N  |q|$) is the spherical Bessel function of the first kind, with
\begin{eqnarray}
j_1 (x) = \frac{\sin x}{x^2} - \frac{\cos x}{x} ~  .
\end{eqnarray}
$r_N$ is the effective nuclear radius, with
\begin{eqnarray}
r_N = \sqrt{ c_A^2 + \frac{7}{3} \pi^2 a^2 - 5 s_\mathrm{skin}^2}   ~  ,
\end{eqnarray}
where $c_A^{} =$ 1.23 $A^{1/3} -$  0.6  fm, and $a =$ 0.52 fm.

Now, for target nuclei with multiple species, the factor $\overline{F}_\mathrm{med}$ is
\begin{eqnarray}
\overline{F}_\mathrm{med}  =
\frac{\sum_i f_i  \int_{E_R^\mathrm{thr}}^{E_{R,i}^\mathrm{high}}  d E_R ~ \epsilon_i(E_R) \frac{d R_i }{d E_R} }
{\sum_i f_i  \int_{E_R^\mathrm{thr}}^{E_{R,i}^\mathrm{high}} d E_R ~ \epsilon_i(E_R) \frac{d R_i }{d E_R}|_{F_{\mathrm{med}}(q^2) = 1}}   ~  ,  \nonumber   \\
\end{eqnarray}
where $f_i$ is the mass fraction of nuclear species $i$ in the detector, and $E_R^\mathrm{thr}$ is the recoil energy threshold of the target nucleus in detections. For a nuclear species $i$: $E_{R,i}^\mathrm{high}$ is the upper boundary of the recoil energy for a given SIDM mass, with
$ E_{R,i}^\mathrm{high}$ being the minimum of the two, $min \big[ 2 \mu_{\Psi N}^2 (v_\mathrm{in}^2)_\mathrm{max} /m_N, E_R^\mathrm{max} \big] $.
$\epsilon_i(E_R)$ is the detection efficiency for a given recoil energy $E_R$. $\frac{d R_i }{d E_R}|_{F_{\mathrm{med}}(q^2) = 1}$ is the differential recoil rate with the factor $F_{\mathrm{med}}(q^2) =$ 1 adopted.


\end{document}